\documentstyle[twocolumn,aps,prl,epsfig]{revtex}

\begin{document}
\draft
\title{Theory for high spin systems with orbital degeneracy}

\twocolumn[\hsize\textwidth\columnwidth\hsize\csname@twocolumnfalse\endcsname
\author{Shun-Qing Shen$^{1,4}$,  X. C. Xie$^{2,5}$, and F. C.  Zhang$^{3}$}
\address{
$^{1}$Department of Physics, The University of Hong Kong, Pokfulam, Hong Kong\\
$^{2}$Department of Physics, Oklahoma State University, Stillwater, OK74078\\
$^{3}$Department of Physics, University of Cincinnati, Cincinnati, Ohio 45221\\
$^{4}$Institute of Physics, Chinese Academy of Sciences, Beijing 100080, China\\
$^{5}$International Center for Quantum Structure, Chinese Academy of Sciences, Beijing, China
}
\date{Nov. 9, 2001}
\maketitle

\begin{abstract}
High-spin systems with orbital degeneracy are studied in the large spin limit. 
In the absence of Hund's coupling, the classical spin model is mapped onto 
disconnected orbital systems with spins up and down, respectively. 
The ground state of the isotropic model is an orbital valence bond state where 
each bond is an orbital singlet with parallel spins, and neighbouring bonds 
interact antiferromagnetically.  The possible relevance to the transition metal oxides are 
discussed.
\end{abstract}


]{\bf \ \ }

In many transition metal oxides, there is an orbital degeneracy in the
electron occupation energy, resulting in rich and novel magnetic phenomena. 
\cite{Tokura00} The orbital ordering and orbital density wave have been
observed experimentally in a family of manganites.\cite{Saitoh01} In these
systems, the spin coupling depends on the electron's orbital occupations.
The simplest model to describe spin-$\frac{1}{2}$ systems with two-fold
orbital degeneracy is the SU(4) model where the Hund's rule coupling and the
orbital anisotropy are neglected.\cite{Li98,Castellani78}\ \ Some transition
metal oxides, such as manganites and lanthanum vanadium oxides, have higher
spins, which has attracted much theoretical interests. \cite
{Kugel80,Ishihara96,Feiner97,Shen00,Khaliulin01} \ Mathematical models for
high spin systems are generally complicated. The Hund's rule coupling favors
ferromagnetic (FM) spins, and the anisotropy of the electron hopping
integrals breaks orbital SU(2) symmetry. While the phase diagrams of these
complex spin systems are rich, there is lack of a tractable method to
systematically study these systems. In this paper, we study a class of spin
systems with two-fold orbital degeneracy in the limits of large spin. Our
model can be derived from a lattice of ions with high spins
ferromagnetically coupled to the strongly correlated electrons with double
orbital degeneracy. In the classical limit of spin, such a system can be
mapped onto two decoupled black and white sub-systems, representing
classical spin up and down, respectively.\ This provides an efficient method
to study the ground states.\cite{method} We apply this method to study the
case where the electron hopping integrals are isotropic and use the
linearized spin wave theory to examine the effects of the quantum spin
fluctuation. The ground state in 1D chain is a spin disordered valence bond
state, in which each bond is an orbital valence bond (orbital singlet,
parallel spins) and neighboring bonds interact antiferromagnetically. The
ground states in a 2D square lattice and in a 3D cubic lattice are spin
antiferromagnetically ordered orbital valence bond (OVB) state. We discuss
possible relevance of the theory to some transition metal oxides.

We start with an Hamiltonian describing correlated electrons coupled to the
spins of local ions, 
\begin{eqnarray}
H &=&t\sum_{\left\langle ij\right\rangle ,\alpha ,\sigma }\left( c_{i\alpha
\sigma }^{\dagger }c_{j\alpha \sigma }+c_{j\alpha \sigma }^{\dagger
}c_{i\alpha \sigma }\right)  \nonumber \\
&&+\text{ }U\sum_{i,\alpha \sigma \neq \alpha ^{^{\prime }}\sigma ^{\prime
}}n_{i\alpha \sigma }n_{i\alpha ^{\prime }\sigma ^{\prime }}-\sum_{i,\alpha
}J_{H}{\bf S}_{i}^{(e)}\cdot {\bf S}_{i}^{(ion)}
\end{eqnarray}
In the above equations, $\alpha =1,2$ are the orbital indices and $\sigma $
is the electron spin. The sum over $\left\langle ij\right\rangle $ runs all
the nearest neighbor (n.n.) pairs. The electronic part of the Hamiltonian is
a generalized Hubbard Hamiltonian with two-fold orbital degeneracy, where
the Hund's rule coupling and anisotropy of the electron hopping integrals
are neglected. ${\bf S}_{i}^{(e)}$ and ${\bf S}_{i}^{(ion)}$ are the spin
operators of electron and ion at site i, respectively. We consider $%
J_{H}\geq 0$, consistent with the Hund's rule. In the strong coupling limit, 
$U\gg t,sJ_{H}$, with $s$ the quantum number of the ion spin, the projection
perturbation theory may be applied to study low energy physics of the
system. At the filling of one electron per site, and up to the order of $%
t^{2}/U$, this leads to an effective Hamiltonian, \cite{Rice95} 
\begin{eqnarray}
H_{p} &=&\frac{J}{2}\sum_{\left\langle i,j\right\rangle }\left( 2{\bf S}%
_{i}^{(e)}\cdot {\bf S}_{j}^{(e)}+1/2\right) \left( 2{\bf T}_{i}\cdot {\bf T}%
_{j}+1/2\right)  \nonumber \\
&&-\sum_{i}J_{H}{\bf S}_{i}^{(e)}\cdot {\bf S}_{i}^{(ion)}
\end{eqnarray}
where ${\bf T}_{i}$ is the electron orbital operator, $J=4t^{2}/U$.
Furthermore, in the case $sJ_{H}>t^{2}/U$, the electron and ion spins
strongly bind ferromagnetically to form a state with total spin $s_{t}=s+1/2$%
. The effective Hamiltonian within this Hilbert space of total spin $s_{t}$
can be obtained by applying the projection operator,\cite{Kubo72} 
\begin{equation}
P=\prod_{i}\frac{2{\bf S}_{i}^{(e)}\cdot {\bf S}_{i}^{(ion)}+(s+1)}{2s+1}.
\end{equation}
The new coupled spin-orbital model is found to be , 
\begin{eqnarray}
H_{eff} &\equiv &PH_{p}P  \nonumber \\
&=&\frac{J}{2s_{t}^{2}}\sum_{\left\langle ij\right\rangle }\left( {\bf S}%
_{i}\cdot {\bf S}_{j}+s_{t}^{2}\right) \left( {\bf T}_{i}\cdot {\bf T}%
_{j}+1/4\right)  \label{model}
\end{eqnarray}
with ${\bf S}_{i}={\bf S}_{i}^{(e)}+{\bf S}_{i}^{(ion)}$ the total spin
operator with maximal eigenvalue $s_{t}$ at site i.

The model has SU(2)$\times $SU(2) symmetry, representing rotational
invariance in both spin and orbital spaces. A special case is at $s_{t}=1/2$%
, corresponding to ${\bf S}_{i}^{(ion)}=0$. In that case, the Hamiltonian
possesses a higher SU(4) symmetry and the model has been studied
extensively. In this paper, we focus on another limit where $s_{t}\gg 1$. We
start with the classical spin to replace ${\bf S}_{i}$ by a classical vector
of length $s_{t}$ described by two angles $\theta _{i}$ and $\phi _{i}$. The
Hamiltonian then reads,\ 
\begin{equation}
H_{c}=J\ \sum_{\left\langle ij\right\rangle }{\bf \cos }^{2}\frac{\Theta
_{ij}}{2}\left( {\bf T}_{i}\cdot {\bf T}_{j}+1/4\right)  \label{Ising}
\end{equation}
where $\Theta _{ij}$ is the angle between the two spin vectors$.$ To
investigate the ground state, we use the variational principle to find the
equations for ${\bf \theta }_{i}$ and $\phi _{i}$: $\ \delta H_{c}/\delta
\theta _{i}=0$ and $\delta H_{c}/\delta \phi _{i}=0.$ We see that $\phi
_{i}=\phi _{j}$ and ${\bf \theta }_{i}={\bf \theta }_{0}$ or/and ${\bf %
\theta }_{0}+\pi $ at all the sites are solutions of the equations. Below we
will consider these solutions. In 1D it can be shown that these solutions
give the lowest energy~\cite{noteexact}. In 2D or 3D we speculate these
solutions contain the lowest energy states, although a rigorous proof is
absent. We set ${\bf \theta }_{0}=0$ below, and label all the lattice sites
with ${\bf \theta }_{i}=0$ by blacks, and all the sites with ${\bf \theta }%
_{i}=\pi $ by whites. The bond Hamiltonian is then reduced to

\begin{equation}
H_{c}(ij)=\left\{ 
\begin{array}{ll}
J({\bf T}_{i}\cdot {\bf T}_{j}+1/4), & \text{if }{\bf \theta }_{i}=\theta
_{j}; \\ 
0, & \text{if }{\bf \theta }_{i}=\theta _{j}\pm \pi .
\end{array}
\right.
\end{equation}
Therefore, the coupled spin-orbital system is decomposed into disconnected
black and white sites or blocks (collection of the connected same colored
sites). In the same colored block, all the spins are parallel but the
orbitals interact antiferromagnetically; and the interaction vanishes
between different blocks. The total energy of the system is then the simple
sum of the these blocks. This greatly simplifies the calculations.

Let us first consider a two-site problem. The ground state of Eq.(\ref{Ising}%
) is an orbital singlet with total spin $2s_{t}$. \ We shall call this
2-site state as an orbital valence bond (OVB), whose energy is $-0.5J$.
There are two competing terms in Eq.(\ref{Ising}). One is the spin-orbital
coupled term, ${\bf \cos }^{2}(\Theta _{ij}/2){\bf T}_{i}\cdot {\bf T}_{j}$,
which favors FM spins and antiferromagnetic (AFM) orbitals. The other is the
spin interaction term ${\bf \cos }^{2}(\Theta _{ij}/2)/4$, which favors AFM
spins. Because of this competition, the ground state of more than 2-sites is
generally not a uniform FM spin state as we will see explicitly below. In
the 1D chain, we divide the chain into blacks and whites and calculate the
lowest energies of these spin configurations. In particular, we consider the
configurations with alternating black and white segments of n-sites (see
Fig. 1a). The total energy of the chain is the sum of these independent
segments. The ground state energies of these segments are calculated by
using exact numerical diagonalization method from $n=2$ to $16$. The results
are plotted in Fig. 2. The energies of the even n segments are lower than
those of the odd n ones. The ground state is an OVB solid with alternating
spins up and down with an energy $-0.25J$ per bond, much lower than that of
the uniform FM spin state corresponding to $n\rightarrow \infty $. The
energy of the latter state can be deduced from the result of the Bethe
ansatz solution, and it is $(1-2\ln 2)J/2=-0.193J$ per bond. Our numerical
results approach to the exact result rapidly as n increases. Including the
spin quantum fluctuation, we expect that the 1D AFM spin long range order be
destroyed based on the dimensionality consideration and the calculation of
the linearized spin wave theory. The system is then described by a spin-$%
2s_{t}$ chain of the OVBs with AFM Heisenberg coupling $J/(32s_{t}^{2})$.
Since $2s_{t}$ is an integer, the ground state is a Haldane's gap state. The
quantum ground state has two-fold degeneracy corresponding to the
translational invariance by one lattice constant in the OVBs.

In a 2D square lattice, we consider various spin configurations including
alternating colored 2-site bond state (Fig.1b), alternating 4-site plaquette
state (Fig.1c), stripes and the uniform FM states. The ground states are
found to be highly degenerate.\cite{Note} \ The lowest energy state in
Fig.1b is an OVB state, and the lowest energy state in Fig. 1c is a
plaquette orbital singlet state whose orbital is given by \cite{Ueda96} {\bf %
t}$(12)${\bf t}$(34)-${\bf t}$(14)${\bf t}$(23)$, with {\bf t}(12)
representing an orbital singlet of sites 1 and 2. The OVB, the plaquette
orbital, and their mixed states are degenerate, and have energy $-0.125J$
per bond. \ The uniform FM spin state has a much high energy $<T_{i}\cdot
T_{j}+1/4>_{2D}=-0.085J$ per bond as estimated from the known result of 2D
spin-1/2 model \cite{Liang90}. The OVB state is also found to have the
lowest energy in 3D cubic lattice. The large degeneracy of the ground states
found in the classical spin limit in higher dimension is removed when the
spin quantum fluctuation is included. \ We have used the spin wave theory to
calculate the energy correction to the classical spin states in 2D square
lattice, and found it to be $-0.01157J/s_{t}$ for the OVB state, and $%
-0.01012J/s_{t}$ for the plaquette orbital state. Therefore the quantum spin
fluctuation favors the OVB state. In 2D and 3D, the AFM spin long range
order is expected to survive from quantum fluctuation. We thus conclude that
the ground state of Eq.(\ref{model}) in 2D and 3D is the OVB state with AFM
ordered spins between the n.n. bonds.

We now turn to discuss the possible relevance of our theory to some
transition metal oxides. We first examine the ground state of cubic vanadate
LaVO$_{3}$. The observed magnetic order of the vanadate is of C-type AFM
phase (FM along the c-axis and AFM in the a-b plane), arising from a
practically undistorted structure above the N\'{e}el temperatures.\cite
{Mahajan92,Nguyen95} To explain the unusual magnetic ordering, Khaliullin 
{\it et al.}\cite{Khaliulin01} recently proposed a spin-orbital Hamiltonian
for the oxides. In the limit $\eta =J_{H}/U\rightarrow 0$ (the notations are
the same as in their paper) and along a given cubic axis, their model is
reduced to Eq.(\ref{model}) in the present paper with $s_{t}=1$. The
corresponding two equivalent orbitals for the coupling along c-axis are, d$%
_{zx}$ and d$_{yz}$, for example. Khaliullin {\it et al.} compared the
ground state energies of the C-type and G-type (AFM in all three directions)
AFM phases. They concluded that the C-type AFM is more stable at $\eta =0$,
and is further stabilized at $\eta >0.$ In that work, the comparison between
the C-type AFM and the OVB state was not included. As we discussed earlier,
the ground state is an OVB state at $\eta =0$ in the large spin limit. We
believe the experimentally observed C-type AFM in that compound is
stabilized by the Hund's coupling. Indeed we have compared the energies of
C-type AFM and OVB state in their model (Eq.(1) of Ref.\cite{Khaliulin01}),
and found that the C-type AFM has lower energy for $\eta >\eta _{c}$ with $%
\eta _{c}\approx 0.06$.~\cite{private} It will be interesting to examine the
higher order spin fluctuation or to use numerical techniques to verify if
the large spin limit applies to spin-1 system.

In the second example, we consider the spin-orbital model for LaMnO$_{3}$
assuming the lattice is not distorted. This type of model has been studied
by many authors and the ground state is A-type AFM (AFM along c-axis and FM
within a-b planes).\cite{Rodriguez98} Here we wish to point out that the
A-type ordering may be obtained by treating the Hund's rule coupling as a
small perturbations. The effective Hamiltonian describing spin $S=2$ and the
two degenerate e$_{g}$ orbitals of the Mn-ion is given in the limit $%
J_{H}\rightarrow 0$,\cite{Kugel80,Ishihara96,Feiner97,Shen00} 
\begin{equation}
H=\frac{J}{2s_{t}^{2}}\sum_{\left\langle ij\right\rangle }\left( {\bf S}%
_{i}\cdot {\bf S}_{j}+s_{t}^{2}\right) \left( \tau _{i}^{\gamma }+\frac{1}{2}%
\right) \left( \tau _{j}^{\gamma }+\frac{1}{2}\right)
\end{equation}
where $\gamma $ is along $j-i$ and the orbital operators $\tau _{i}^{\gamma
}=\cos (2m_{\gamma }\pi /3){\bf T}_{i}^{z}-\sin (2m_{\gamma }\pi /3){\bf T}%
_{i}^{x}$ ($m_{\gamma }=1,2,3$) with eigenvalues $\pm \frac{1}{2}$. The
ground state of this Hamiltonian in the large spin limit is highly
degenerate. As the expectation value of $\left( \tau _{i}^{\gamma }+\frac{1}{%
2}\right) \left( \tau _{j}^{\gamma }+\frac{1}{2}\right) $ is always not less
than zero, the low bound for the ground state energy is zero. Thus the G-,
C-, and A-type AFM are all degenerate ground states with zero energy. When a
small Hund's rule coupling is introduced, the high degeneracy in the ground
state is removed. To simplify our problem, we keep the perturbation term due
to the Hund's rule coupling up to the order $JJ_{H}/U$, $\Delta
H=-(J/2s_{t}^{2})\sum \left( {\bf S}_{i}\cdot {\bf S}_{j}+s_{t}^{2}\right) h$%
, where $h$ is of order $sJ_{H}/U\ll 1$. Including $\Delta H$, the A-type
AFM spin state is most favorable in energy. In that state, the orbitals are
C-type antiferro-orbitally ordered: d$_{x^{2}-z^{2}}$/d$_{y^{2}-z^{2}}$
alternative in the a-b plane, and ferro-orbital along the c-axis. This state
is favored because of the energy gain of $\Delta H$ from the bonds within
the a-b planes. Our calculation shows that the superexchange interaction
including the Hund's coupling leads to a spin A-type and orbital C-type
structure. This state might be realized in KCuF$_{3}.$\cite{Kugel80} Note
that the orbital ordering pattern is different from that experimentally
observed in LaMnO$_{3}$ with a d$_{3x^{2}-r^{2}}$/d$_{3y^{2}-r^{2}}$
alternating ordering. The observed orbital ordering is due to strong
Jahn-Teller distortion \cite{Allen99}, but not due to the superexchange \cite
{Tokura00}.

Finally, we examine an extended model of Eq.(\ref{model}) in a cubic lattice
to include the anisotropic coupling in different axis and the Hund's rule
coupling, 
\begin{equation}
H=\sum_{i,a}J_{\alpha }\left( {\bf S}_{i}\cdot {\bf S}_{i+a}+s_{t}^{2}%
\right) ({\bf \ T}_{i}\cdot {\bf T}_{j}+\frac{B}{4})
\end{equation}
where $J_{\alpha }$ is the coupling strength along the direction $\alpha
=a,b,c$. At the symmetric point $B=1$ and $J_{a}=J_{b}=J_{c},$ the model is
reduced to Eq.(\ref{model}) . In the presence of $J_{H},$ $B<1$ (see in $%
\Delta H$ in the previous example). We compare the ground state energies of
various states including OVB state, the uniform FM spin state, and the A-,
C-, and G-type AFM. In Fig. 3, we plot the phase diagram for the lowest
energy states in the parameter space B and $\eta =J_{ab}/J_{c}.$ The OVB\
state is found to be stable ground state in a finite parameter space around
the symmetry point. It will be interesting to find an experimental
realization of the OVB state. Since the OVB state break the translational
symmetry, we expect the electron-lattice interaction may accompany a crystal
structure phase transition to result in shorter and longer bonds. At a
larger $\eta $ and smaller $B$, C-type AFM spin state is realized, in which
the orbitals are disordered liquid. At a smaller $\eta $ and smaller $B$,
A-type AFM\ becomes stable, in which the orbitals has a long range AFM order
in the x-y plane, and FM along the z-axis. The phase diagram is thus very
rich, reminiscent of some of the features in the transition metal oxides.

In conclusion, we have examined the ground state of a coupled high-spin
orbital model from the large spin limit point of view. A novel orbital
valence bond state is found to be stable in a certain parameter region, and
may be realized in future experiments. The anisotropic model contains both
C-type and A-type AFM spin states among others. Our method should be useful
in study of more realistic models when Hund' rule coupling is weak. The
large spin approach emphasizes orbital quantum fluctuation over spins, and
special care is needed to apply the method to spin-1/2 case. In the SU(4)
limit, there is a permutation symmetry between spin and orbital, and we do
not expect the method will work. The quantum fluctuation may drive the two
degenerated spin and orbital valence bond states to a quantum state with
more intricate correlations. In fact, from the Bethe ansatz solution, the 1D
chain has a gapless SU(4) liquid ground state, which is not dimerized.\cite
{Sutherland75}

We would like to thank Jian Wang and Michael Ma for useful discussions. This
work was in part supported by the RGC Grant No. HKU 7088/01P from Hong Kong,
and by the US DOE Grant No. DE/FG03-01ER45687, and by the Chinese Academy of
Sciences.

\begin{figure}[tbp]
\caption{Illustration of several possible lowest energy states for the
model. The solid line represents OVB with spin up, and dashed line
represents OVB with spin down. (a). A one-dimensional OVB solid. (b). A
two-dimensional OVB\ solid. (c) A two-dimensional plaquette RVB solid.}
\end{figure}

\begin{figure}[tbp]
\caption{Energy per bond ($Js_{t}^{2}/2$) versus the site number per cluster
on a one-dimensional chain.The solid line corresponds to the exact energy of
one-dimensional chain by means of Bethe ansatz.}
\end{figure}

\begin{figure}[tbp]
\caption{The phase diagram for high spin systems with orbital degeneracy in
Eq. (5). OVB(ab) represents the OVBs aligning along the a-or b-axis, and
OVB(c) represents the OVBs aligning along the c-axis.}
\end{figure}


\begin{references}
\bibitem{Tokura00}  Y. Tokura and N. Nagaosa, Science 288, 462 (2000).

\bibitem{Saitoh01}  E. Saitoh et al, Nature (London) 410, 180 (2001).

\bibitem{Li98}  Y. Q. Li {\it et al.}, Phys. Rev. Lett. 81, 3527 (1998).

\bibitem{Castellani78}  C. Castellani {\it et al.}, Phys. Rev. B 18, 4945
(1978).

\bibitem{Kugel80}  K. I. Kugel' and D. I. Khomskii, Fiz. Nizk. Temp. 6, 207
(1980) (Sov. J. Low Temp. 6, 99 (1980)).

\bibitem{Ishihara96}  S. Ishihara {\it et al.}, Physica C 263, 130 (1996).

\bibitem{Feiner97}  L. F. Feiner {\it et al.}, Phys. Rev. Lett. 78, 2799
(1997).

\bibitem{Shen00}  S. Q. Shen and Z. D. Wang, Phys. Rev. B 61, 9532 (2000).

\bibitem{Khaliulin01}  G. Khaliullin {\it et al.}, Phys. Rev. Lett. 86, 3879
(2001).

\bibitem{method}  This method was implicitly used by some authors.

\bibitem{Rice95}  T. M. Rice, Spectroscopy of Mott Insulators and Correlated
Metals, edited by A. Fujimori and Y. Tokura (Springer, Berlin, 1995).

\bibitem{Kubo72}  K. Kubo and N. Ohata, J. Phys. Soc. Jpn 33, 21 (1972); S.
Q. Shen and Z. D. Wang, Phys. Rev. B58, R8877 (1998).

\bibitem{noteexact}  In 1D and for an arbitrary orbital configuration, we
can assign a spin configuration in which two neighboring spins are parallel
if $(\vec T_i \cdot \vec T_j +1/4)$ is negative, and are antiparallel
otherwise. Such a spin configuration gives the lowest energy for each bond
in the corresponding orbital state. Therefore, the solutions we consider
contain the lowest energy states.

\bibitem{Note}  We have evaluated the energies for a series of clusters
consisting of up to eighteen sites by exact diagonalization. Similar to the
1D chain, larger-size clusters tend to have higher energies than a
collection of smaller-size clusters with the same number of lattice sites.
For example, the energy for a $2\times 4$ cluster is higher than the energy
for two $2\times 2$ or four $1\times 2$ clusters.

\bibitem{Ueda96}  K. Ueda {\it et al.}, Phys. Rev. Lett. 76, 1932 (1996).

\bibitem{Liang90}  S. D. Liang, Phys. Rev. 42, 6555 (1990).

\bibitem{Mahajan92}  A. V. Mahajan {\it et al.}, Phys. Rev. B 46, 10 966
(1992).

\bibitem{Nguyen95}  H. C. Nguyen {\it et al.}, Phys. Rev. B 52, 324 (1995).

\bibitem{private}  After we completed the work, we learned that the valence
bond state was recently also considered by Khaliullin et al. in a paper in
preparation, private communication.

\bibitem{Rodriguez98}  J. Rodriguez-Carvajal {\it et al.}, Phys. Rev. B 57,
R3189 (1998). Also see Y. Tokura (ed.), Colossal Magnetoresistive Oxides,
(Gorden and Breach, 2000).

\bibitem{Allen99}  P. B. Allen {\it et al.}, Phys. Rev. B60, 10747 (1999).

\bibitem{Sutherland75}  B. Sutherland, Phys. Rev. B 12, 3795 (1975).
\end{references}
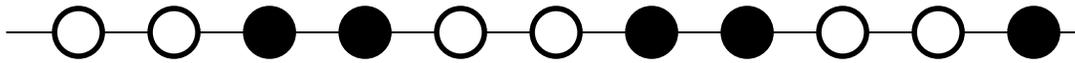
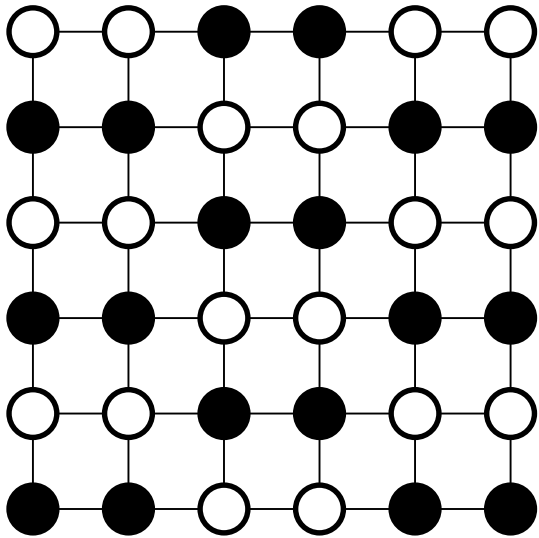
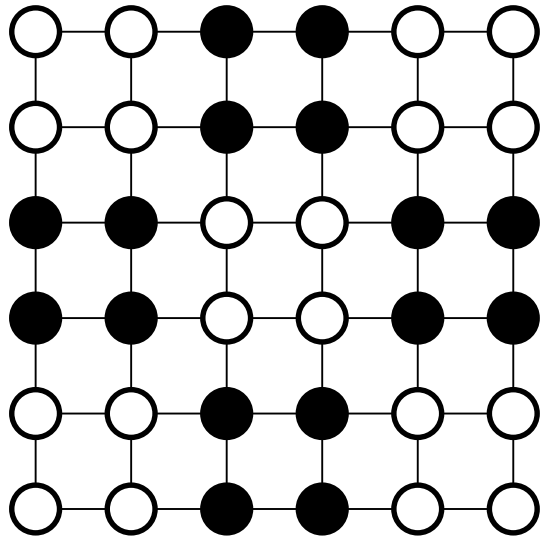
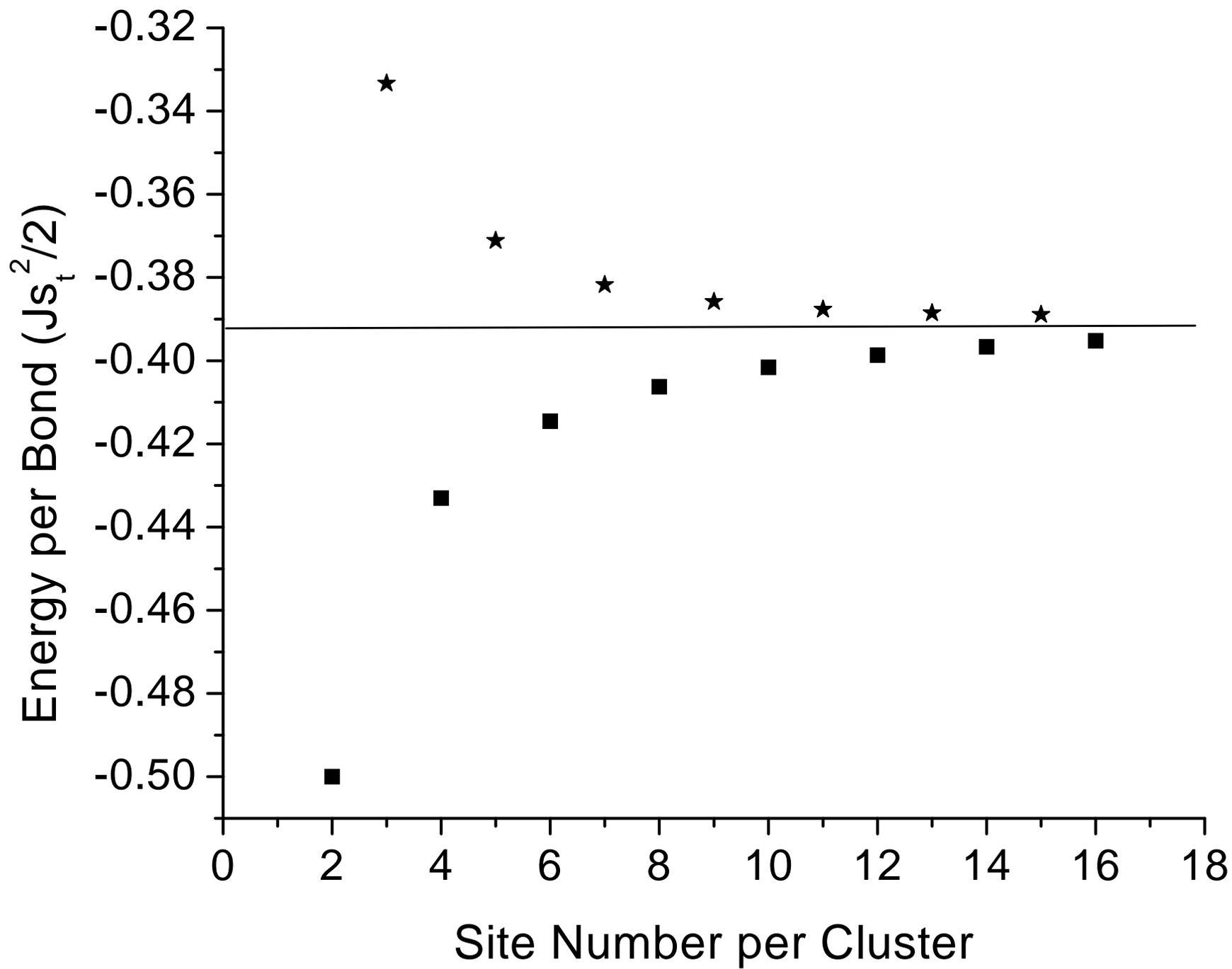
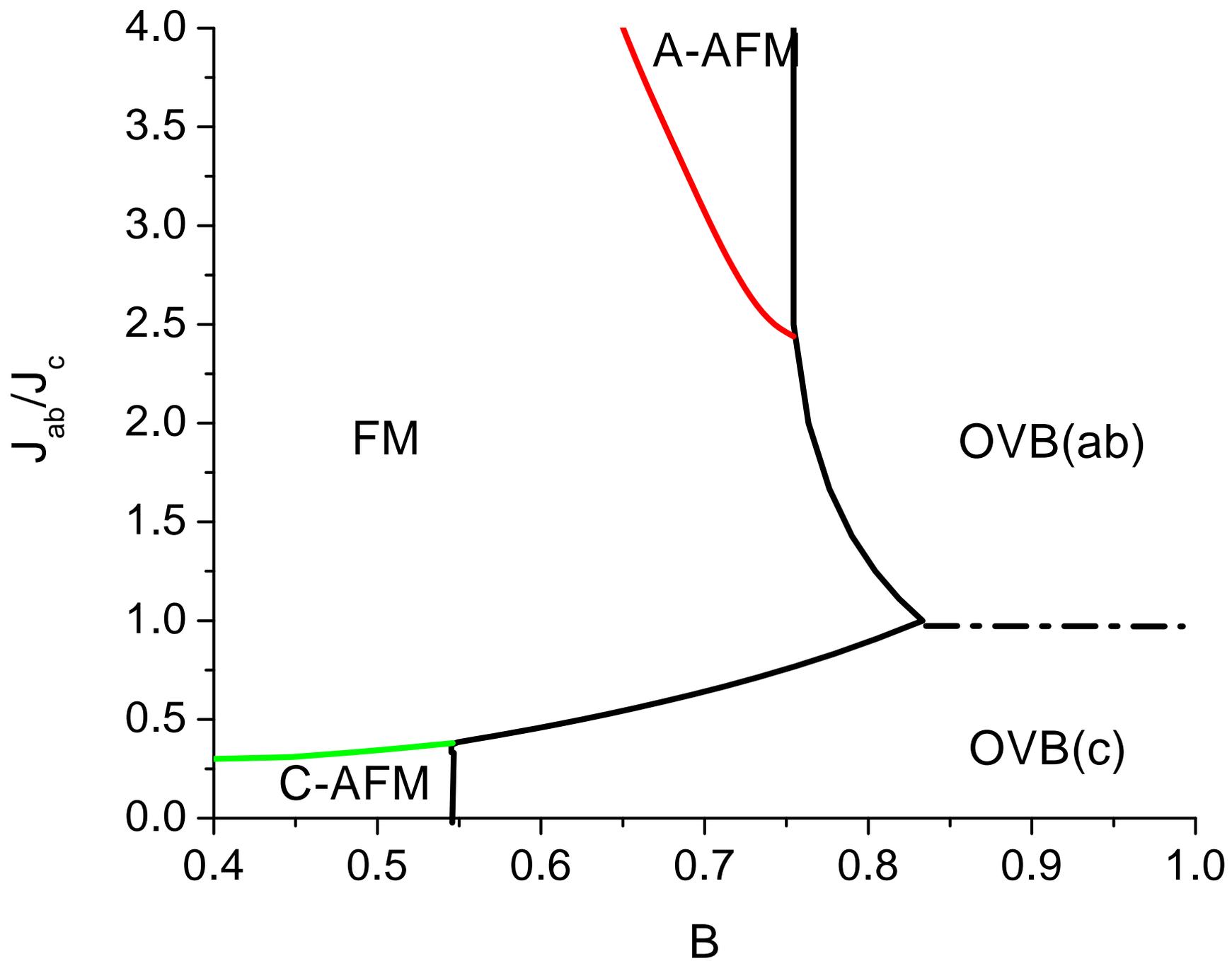
\end{document}